\documentclass[hyper]{JHEP3}


\skip\footins = 1\bigskipamount plus 2pt minus 4pt

\usepackage{epsfig}
\newcommand{\pv}[1]{{-  \hspace {-4.0mm} #1}}

\title{Solitons in the duality-based matrix model}

\author{V. Bardek$^{\dagger}$ and S. Meljanac$^{\dagger}$\\

$^{\dagger}$
Institute Rudjer Bo\v skovi\'c, Bijeni\v cka cesta 54,\\
P.O. Box 180, HR-10002 Zagreb, Croatia\\
E-mail: \email{bardek@irb.hr},
\email{meljanac@irb.hr}}

\abstract{We analyze
soliton solutions in the duality-based matrix model.
There are two types of solution, a one soliton-antisoliton solution
(with the constant boundary condition at infinity) and a periodic solution
with an infinite number of solitons. It is shown that there is no finite
number $ \;  (n > 1 ) \; $ of solitons at finite distances in the
limit when the length of the box tends to infinity. Particularly,
there is no finite number of $ \; \delta - \; $ function solitons in the singular limit.}

\begin{document}


\section{Introduction}

In \cite{Andric:2004dv}, the ``duality-based matrix model'' was
proposed in the collective-field formulation.It was conjectured to
describe the hermitean matrix model. In the next paper \cite{Andric:2004cb},
the same authors studied solitons and excitations in the duality-
based matrix model. They claimed that there existed many soliton
solutions with the constant boundary condition at infinity, i.e.
topologically nontrivial BPS solitons. Furthermore, in the paper \cite{Andric:2006ve}
they constructed solitons and giants in matrix models. The two-field
model, associated with two types of  giant graviton, was proposed as
the duality-based matrix model. They claimed that they found the finite form of
the $ \;  n-  \; $ soliton solution. The singular limit of this solution,
particularly the finite number of $ \; \delta - \; $ function solutions
was discussed.

On the other hand \cite{Bardek:2005yx}, we have studied BPS solitons
in the Calogero model with distinguishable particles in the collective-
field approach. This model includes the duality-based matrix model as
a special case.
 We studied soliton solutions with the constant boundary
condition at infinity. We only found a one-soliton-antisoliton solution
and explicitly showed that there did not exist a finite two-soliton
solution with the constant boundary condition. In the comment
\cite{Bardek:2005cs} we showed explicitly that there did not exist
a singular limit of the form $ \; (1 - \lambda) [\delta(x - a) + \delta(x + a)],  \; $
with $ \; a \; $ finite.

The purpose of this paper is to present an independent analysis of
 soliton solutions in order to confirm the correctness of our previous
 conclusions concerning the existence of only one soliton-antisoliton
 pair of solution in the $ \; L \rightarrow \infty \; $ limit.
 In this paper we extend a general construction \cite{Bardek:2005yx}
to periodic soliton solutions. We also find a periodic infinite number of solitons
with an arbitrary finite period when the length of the box $ \; L \; $ tends
to infinity. However, in contrast to \cite{Andric:2006ve}, we show
that there is no finite number $ \;  (n > 1 ) \; $ of solitons at finite
distances when $ \; L \rightarrow \infty. \; $ Consequently, there is no
finite number of $ \; \delta  \; $ functions in the singular limit.


\section{BPS equations and their solutions}

The Hamiltonian  \cite{Andric:2004dv}, \cite{Andric:2004cb}, \cite{Andric:2006ve},
 \cite{Bardek:2005yx}
 for the duality-based matrix model in the collective-field
formulation is
$$
 H =  \int dx \frac{\rho_{1}(x)}{2m_{1}} \bigg[  { ( \partial_{x} \pi_{1}(x))}^{2}
 + {\left( \frac{\lambda_{1} - 1}{2} \frac{\partial_{x} \rho_{1}}{\rho_{1}} +
  \pv \int dy \frac{\lambda_{1} \rho_{1}(y) + \rho_{2}(y)}{x - y}
 \right)}^{2} \bigg]
$$
\begin{equation} \label{hamiltonian}
 + \int dx \frac{\rho_{2}(x)}{2m_{2}} \bigg[  { ( \partial_{x} \pi_{2}(x))}^{2}
 + {\left( \frac{\lambda_{2} - 1}{2} \frac{\partial_{x} \rho_{2}}{\rho_{2}} +
  \pv \int dy \frac{\lambda_{2} \rho_{2}(y) + \rho_{1}(y)}{x - y}
 \right)}^{2} \bigg],
\end{equation}
where
\begin{equation} \label{threebody}
  \frac{\lambda_{1}}{{m_{1}}^{2}} = \frac{\lambda_{2}}{{m_{2}}^{2}} =
   \frac{1}{m_{1} m_{2}},  \quad  \lambda_{1} \lambda_{2} = 1, \quad
          0 < \lambda_{1} < 1.
\end{equation}
The corresponding BPS equations
   \cite{Andric:2004cb}, \cite{Andric:2006ve},
 \cite{Bardek:2005yx}, \cite{Bardek:2005cs} are given by
$$
({\lambda}_{1} - 1) \partial_{x} {\rho}_{1} = 2 \pi {\rho}_{1}
 ({\lambda}_{1} {\rho}_{1}^{H} + {\rho}_{2}^{H}),
$$
\begin{equation} \label{bps}
  ({\lambda}_{2} - 1) \partial_{x} {\rho}_{2} = 2 \pi {\rho}_{2}
 ({\lambda}_{2} {\rho}_{2}^{H} + {\rho}_{1}^{H}),
\end{equation}
where
\begin{equation} \label{rhoh}
  {\rho}^{H}(x) = \frac{1}{\pi} \pv \int \frac{ dy \rho(y)}{ y - x}.
\end{equation}
Then it follows
\begin{equation} \label{solution}
  {\rho}_{1} {\rho}_{2} = c,  \quad  {\lambda}_{i}  \neq 1.
\end{equation}
In this paper, we extend our construction of a one soliton-antisoliton solution
\cite{Bardek:2005yx} to periodic solutions of
 \cite{Polychronakos:1994xg}, \cite{Sen:1997qt} and obtain a general
solution of two coupled BPS equations in the form presented in \cite{Bardek:2005yx}. For the
solutions with the constant boundary condition at infinity
 $ \;( {\rho}_{1}, {\rho}_{2} \rightarrow \mbox{const}),  \;\;
 \mbox{for} \;\;  |x| \rightarrow \infty, \;$
we construct a one soliton-antisoliton solution \cite{Bardek:2005yx},
$$
 {\rho}_{1}(x) = \alpha + r_{1}(x) > 0,
$$
and
\begin{equation} \label{solution1}
  {\rho}_{2}(x) = \frac{c}{\alpha} + r_{2}(x) > 0,
\end{equation}
where $ \; r_{1}(x), r_{2}(x)  \rightarrow 0 \; $ when $ \; |x| \rightarrow \infty. \;$
The functions $ \; r_{1}(x), r_{2}(x)  \;$ satisfy
$$
({\lambda}_{1} - 1) \partial_{x} r_{1} = 2 \pi {\lambda}_{1}r_{1}
r_{1}^{H},   
$$
\begin{equation} \label{bps1}
  ({\lambda}_{2} - 1) \partial_{x} r_{2} = 2 \pi {\lambda}_{2} r_{2}
  r_{2}^{H},  
\end{equation}
and are given by
\begin{equation} \label{firstfamily}
  r_{1}(x) = \frac{{\lambda}_{1} - 1}{ \pi {\lambda}_{1}} \frac{b}{x^{2} + b^{2}},
\end{equation}
\begin{equation} \label{secondfamily}
 r_{2}(x) =
 \frac{{\lambda}_{2} - 1}{ \pi {\lambda}_{2}} \frac{a}{x^{2} + a^{2}},
\end{equation}
where $ \; a, b > 0. \;$
From Eq. (\ref{solution}) and the normalization conditions
\begin{equation} \label{integ1}
  \int dx {\rho}_{1}(x) = N_{1} = L \alpha - \frac{1- {\lambda}_{1}}{{\lambda}_{1}},
\end{equation}
\begin{equation} \label{integ2}
  \int dx {\rho}_{2}(x) = N_{2} = L \frac{c}{\alpha} + 1- {\lambda}_{1},
\end{equation}
we find $ \; (N_{1}, N_{2} >> 1) \;$
$$
 a = \frac {N_{1} ( {\lambda}_{1} -1)}{ \alpha \pi N_{2}
 ( 1 -  {N_{1}}^{2} { {\lambda}_{1}}^{2} / {N_{2}}^{2} )},
$$
$$
 b = \frac { (1 -  {\lambda}_{1})}{{\lambda}_{1} \alpha \pi
  ( 1 -  {N_{2}}^{2} /  {{\lambda}_{1}}^{2}{N_{1}}^{2} )},
$$
\begin{equation} \label{abc}
 c = {\alpha}^{2} \frac{N_{2}}{N_{1}}.
\end{equation}
We note in passing that for $ \; N_{2} = {\lambda}_{1} N_{1}, \;$ it follows $ \; a,b =
\infty, \; {\rho}_{1}(x) = \alpha, \; {\rho}_{2}(x) = \frac{c}{\alpha} =
{\lambda}_{1} \alpha.  \;$

We point out that our construction \cite{Bardek:2005yx} for
$ \; {\rho}_{1}(x) = \alpha + r_{1}(x), \;\;  {\rho}_{2}(x) = \frac{c}{\alpha} + r_{2}(x), \;$
(Eq. (\ref{solution1})), automatically satisfies the
coupled BPS equations of a two-family system, (Eqs.(\ref{bps})).
To see that this is indeed the case, one has to use  Eqs. (\ref{solution}) and (\ref{solution1}) and
rewrite  $ \; {\rho}_{1} \; $ and $ \; {\rho}_{2} \; $ in terms of
  $ \; r_{1} \; $ and $ \; r_{2} \; $ as
 $ \; {\rho}_{1} = \frac{ c}{\alpha} \frac{r_{1}}{r_{2}} \; $
and $ \; {\rho}_{2} = \alpha \frac{r_{2}}{r_{1}}. \; $
 By differentiating these relations with respect to $ \; x \; $
and using Eqs.(\ref{bps1}), we end up with the coupled BPS equations
(\ref{bps}).
In this way, we have reduced the two-family Calogero model to two
one-family Calogero decoupled systems (\ref{bps1}). We note that the
solutions of the BPS equation are simultaneously the solutions of the
corresponding variational equation.
 The problem of finding solutions of the variational
equation corresponding to a one-family Calogero model was solved in
\cite{Polychronakos:1994xg}, \cite{Sen:1997qt} a long time ago. From these solutions
and by using the identities
$$
 {\left (  \frac{b}{x^{2} + b^{2}}   \right )}^{H}
    =  \frac{x}{x^{2} + b^{2}}
$$
and
\begin{equation} \label{identities}
 {\left ( \frac{\sinh u}{\cosh u -  \cos kx}   \right )}^{H}
    =  \frac{\sin kx}{\cosh u -  \cos kx},
\end{equation}
one easily finds only two types of solutions of the BPS equations (\ref{bps1}).
\begin{enumerate}
\item Aperiodic one-soliton solution
 \begin{equation} \label{aper}
  r(x) = \frac{\lambda - 1}{ \pi \lambda} \frac{b}{x^{2} + b^{2}},
  \;\;\;\; b \in R_{+}
 \end{equation}
  with the property
\begin{equation}
  \int_{- \infty}^{\infty} dx r(x) = \frac{\lambda - 1}{\lambda}.
  \end{equation}
\item Periodic  solutions
 \begin{equation} \label{per}
    r(x) = \frac{\lambda - 1}{2 \pi \lambda} k \frac{\sinh u}{\cosh u -
    \cos kx},   \;\;\;  \mbox{where} \;\;\; u \ge 0, \;\; k \in R_{+},
 \end{equation}
  with the property
\begin{equation}
  \int_{- \frac{ \pi}{k}}^{\frac{ \pi}{k}} dx r(x) = \frac{\lambda - 1}{\lambda}.
  \end{equation}
\end{enumerate}
The period $ \; \frac{2 \pi}{k} \;$ is arbitrary. In the limit
$ \; k \rightarrow 0, \;$ we find that the period $ \; \frac{2 \pi}{k} \rightarrow \infty \;$
and, if $ \; u \rightarrow 0, \;$ one obtains the first ``aperiodic solution'' with
finite $ \; b =  \frac{2 \sinh \frac{u}{2}}{k}. \;$
Using our construction for the aperiodic one-soliton solution, we find a
unique one-soliton-antisoliton solution \cite{Bardek:2005yx} and there
is no other finite number $ \;  (n > 1 ) \; $ of soliton solutions (at
mutually finite distances) of Eq. (\ref{bps}).

Let us now apply our general construction, Eqs. (\ref{solution1}), to the
periodic solutions $ \; r_{1}(x), r_{2}(x)  \;$ of
Eqs. (\ref{bps1}) of the form (\ref{per}) with the
same period $ \; \frac{2 \pi}{q}. \;$ We obtain
\begin{equation} \label{r1}
  r_{1}(x) = \frac{{\lambda}_{1} - 1}{2 \pi {\lambda}_{1}} q
  \frac{\sinh u_{1}}{\cosh u_{1} - \cos qx },
\end{equation}
\begin{equation} \label{r2}
 r_{2}(x) =
 \frac{{\lambda}_{2} - 1}{2 \pi {\lambda}_{2}} q \frac{\sinh u_{2}}{\cosh u_{2} - \cos qx },
\end{equation}
where $ \; u_{1} > u_{2} \ge 0  \;$ and $ \; r_{1}(x), r_{2}(x)  \;$ are
related by the relation (\ref{solution}).
The parameters $ \; u_{1}, u_{2}  \;$ can be expressed in terms of
$ \; {\rho}_{10} = \frac{N_{1}}{L}, \;\; {\rho}_{20} = \frac{N_{2}}{L} \;$
and $ \; q. \;$ ($ \; N_{1}, N_{2} \rightarrow \infty, \;\; L
\rightarrow \infty \; $ and $ \; {\rho}_{10}, {\rho}_{20}, q \;$ are
finite). Note that $ \; q \;$ is a free arbitrary parameter which
cannot be determined by the parameters of the model.
From Eq. (\ref{solution}) we find
\begin{equation} \label{i1}
  \alpha = \frac{1 - {\lambda}_{1}}{2 \pi {\lambda}_{1}} q
  \frac{\sinh u_{1}}{\cosh u_{1} - \cosh u_{2} } > 0,
\end{equation}
\begin{equation} \label{i2}
 \frac{c}{\alpha} =
 \frac{1 - {\lambda}_{1}}{2 \pi} q \frac{\sinh u_{2}}{\cosh u_{1} -
 \cosh u_{2} } > 0,
\end{equation}
\begin{equation} \label{i3}
 \frac{c}{{\lambda}_{1} {\alpha}^{2}} =
  \frac{\sinh u_{2}}{\sinh u_{1}},
\end{equation}
whereas from Eqs. (\ref{solution1}) it follows
$$
  \alpha = {\rho}_{10} + \frac{1 - {\lambda}_{1}}{2 \pi {\lambda}_{1}} q,
$$
\begin{equation} \label{i4}
 \frac{c}{\alpha} = {\rho}_{20} -
 \frac{1 - {\lambda}_{1}}{2 \pi} q.
\end{equation}
Using equations (\ref{i1} - \ref{i4}) we obtain
$$
  \coth \frac{u_{1} + u_{2}}{2}
   = 2 + \frac{{\lambda}_{1} {\rho}_{10} - {\rho}_{20} }{(1 - {\lambda}_{1}) q} 2 \pi,
$$
\begin{equation} \label{i5}
 \coth \frac{u_{1} - u_{2}}{2}
   = \frac{{\lambda}_{1} {\rho}_{10} + {\rho}_{20} }{(1 - {\lambda}_{1}) q} 2 \pi.
\end{equation}
Note that for $ \; q \rightarrow 0, \;$ we have $ \; \alpha \rightarrow
{\rho}_{10},  \;$ $ \; \frac{c}{\alpha} \rightarrow{\rho}_{20}, \;$
$ \; u_{1}, u_{2} \rightarrow 0, \;$ (except for $ \; {\lambda}_{1}
{\rho}_{10} = {\rho}_{20} \;$ when $ \; u_{1} = u_{2}, \;$$ \;  \coth u_{1} = 2 ).$
In this limit, we obtain the parameters $ \; a,b,c, \;$ Eqs. (\ref{abc}),
found in \cite{Bardek:2005yx},
$ \; a = \frac{2 \sinh \frac{u_{2}}{2}}{q}, \;\;\; b = \frac{2 \sinh
  \frac{u_{1}}{2}}{q}, \;\;\;  \frac{c}{ {\alpha}^{2}} =
\frac{N_{2}}{N_{1}}. \;$

For $ \; u_{2} \rightarrow 0 \;$ and finite $ \; q \;$ and $ \;  u_{1}, \;$
 there exists an infinite number of $ \; \delta - \;$
function soliton solutions with the finite period $ \; \frac{2 \pi}{q}, \;$
 with $ \; (1 - {\lambda}_{1})q =
2 \pi {\rho}_{20}. \;$ Namely, the soliton solutions are
\begin{equation} \label{i6}
 {\rho}_{1}(x) = \alpha + \frac{{\lambda}_{1} - 1}{2 \pi {\lambda}_{1}} q
  \frac{\sinh u_{1}}{\cosh u_{1} - \cos qx }
   = \alpha \frac{{\sin}^{2} \frac{qx}{2}}{{\sinh}^{2} \frac{u_{1}}{2} + {\sin}^{2} \frac{qx}{2}},
\end{equation}
\begin{equation} \label{i7}
 {\rho}_{2}(x) = (1 - {\lambda}_{1}) \sum_{i \in Z} \delta  ( x -
 \frac{2 \pi}{q} i),
\end{equation}
where $ \; {\rho}_{1}(x) {\rho}_{2}(x) = 0 \;$ and
 $ \; \alpha = \frac{1 - {\lambda}_{1}}{2 \pi {\lambda}_{1}} q \coth
 \frac{u_{1}}{2}. \;$ Note that for $ \;  u_{1} = u_{2} = 0 \;$ and
 $ \; q = 0, \;$ our  solutions
 reduce to $ \; {\rho}_{1} = \alpha \;$ and
$ \; {\rho}_{2} = \frac{c}{\alpha}. \;$

 \section{Discussion and conclusion}

\begin{enumerate}
\item Our general construction of the solutions of the coupled BPS equations (\ref{bps}),
   $ \; {\rho}_{1} = \alpha + r_{1}, \;\;  {\rho}_{2} = \frac{c}{\alpha}
   + r_{2}, \; $ (\ref{solution1}), leads to the decoupled BPS equations (\ref{bps1}) for one-family
   Calogero models for $ \; r_{1}(x), r_{2}(x),  \;$ respectively,
   satisfying the relation $ \; {\rho}_{1} {\rho}_{2} = c, \;$ (\ref{solution}).
\item There is only a one soliton-antisoliton solution ( \ref{firstfamily}),
(\ref{secondfamily}) with the constant boundary
  condition at $\;  |x| \rightarrow \infty, \;$ \cite{Bardek:2005yx}
 when $ \; q \rightarrow 0. \;$ There is also a periodic solution (\ref{r1}),(\ref{r2}) with
 a infinite number of solitons, with the finite period $ \; \frac{2 \pi}{q} \;$
 and $ \; L \rightarrow \infty. \;$
\item However, there is a general result that there are no solutions of the
  BPS Eq. (\ref{bps}) with a finite number $ \;  (n > 1 ) \; $ of solitons,
  mutually at finite distances, when $ \; L \rightarrow \infty. \;$
  Namely, for a finite number of solitons $ \; n  \; $ in the box of length $ \; L, \;$
 $ \; q = \frac{2 \pi n}{L} \rightarrow 0 \;$ when $ \; L \rightarrow \infty. \;$
 Then the period $ \; \frac{2 \pi }{q} = \frac{L }{n} \rightarrow \infty \;$
  and one obtains only a one soliton-antisoliton solution \cite{Bardek:2005yx},
  in contradiction with \cite{Andric:2004cb}, \cite{Andric:2006ve}.
  Let us mention that
  we have explicitly shown that there does not exist a two-soliton
solution at finite distance, with the constant boundary condition at
infinity \cite{Bardek:2005yx}, \cite{Bardek:2005cs} .
\item Consequently, there is no finite number $ \;  (n > 1 ) \; $ of
  $ \; \delta  \;$ functions at mutually finite distances. Namely,
   $ \; q \rightarrow 0 \;$ and
 $ \;  u_{2} \rightarrow 0 \;$ imply $\; {\rho}_{20} \rightarrow 0, \;$
 which is in contradiction with $ \; {\rho}_{10}, {\rho}_{20}  \;$ finite.
\item
Moreover, from the periodic $ \; \delta - \;$ function solution with
finite  $ \; q  \;$
\begin{equation}
 {\rho}_{1}(x) =
    \alpha \frac{{\sin}^{2} \frac{qx}{2}}{{\sinh}^{2} \frac{u_{1}}{2} + {\sin}^{2} \frac{qx}{2}},
\end{equation}
\begin{equation}
 {\rho}_{2}(x) = (1 - {\lambda}_{1}) \sum_{i \in Z} \delta  ( x -
 \frac{2 \pi}{q} i),
\end{equation}
where
 $ \; \alpha = \frac{1 - {\lambda}_{1}}{2 \pi {\lambda}_{1}} q \coth \frac{u_{1}}{2}, \;$
we can easily see that
when we take the limit $ \; q \rightarrow 0, \;$ then, for finite $ \; u_{1}, \;$
 $ \; \alpha \rightarrow 0 \;$
  and $ \; {\rho}_{1}(x) \rightarrow 0, \;$
leading to an unacceptable solution. In order that $ \; \alpha \neq 0, \;$ one
 has to take the limit $ \; u_{1} \rightarrow 0. \;$
When $ \; \frac{u_{1}}{ q} = b = \mbox{finite}, \;$ we obtain only an aperiodic
 singular solution
\begin{equation}
 {\rho}_{1}(x) =
    \alpha  \frac{x^{2}}{x^{2} + b^{2}},
\end{equation}
\begin{equation}
 {\rho}_{2}(x) = (1 - {\lambda}_{1}) \delta (x)
\end{equation}
found in \cite{Bardek:2005yx}.
 Hence, there is no finite
 number $ \;  (n > 1 ) \; $ of $ \; \delta  \;$ functions in the
 solution $ \; {\rho}_{2},  \; $ when $ \; q \rightarrow 0. \;$

Particularly, we again point out \cite{Bardek:2005yx}, \cite{Bardek:2005cs} that
$ \;\;
 {\rho}_{2}(x) = (1 - {\lambda}_{1}) \left( \delta  ( x-a) + \delta  (
 x+a) \right ), \;$ $ \; |a| < \infty, \;\; $
and
$ \;\;
 {\rho}_{1}(x) = {\rho}_{0} \frac{{(x^{2} - a^{2})}^{2}}
{(x^{2} + b^{2})(x^{2} + {\bar{b}}^{2})} ,
 \;\;\; {\rho}_{1}{\rho}_{2} = 0,\;\; $ is not a solution of Eqs. (\ref{bps}),
 \cite{Bardek:2005cs}, in contradiction with \cite{Andric:2004cb}, \cite{Andric:2006ve},
  implying that their conclusion and
 conjecture (rational ansatz) \cite{Andric:2004cb} are wrong.
\item In view of all afore-mentioned findings, we conclude that there is no finite
 $ \; (n > 1) \; $ number of soliton solutions even in the genuine one-family Calogero
 model $ \; (L \rightarrow \infty). \; $
\end{enumerate}

We note that the duality-based matrix model \cite{Andric:2004dv},
\cite{Andric:2004cb}, \cite{Andric:2006ve} is not built on exact
duality between two one-family Calogero models,
\cite{Minahan:1994ce}, \cite{[11]}, \cite{Bardek:2006yt}. The
two-family (multispecies) Calogero models with exact duality
symmetry have been constructed in the collective-field formulation
in a natural and unique way \cite{Bardek:2006yt}. The analysis of
the corresponding soliton solutions is in preparation \cite{prep}.

\bigskip
\bigskip

{\bf Acknowledgment}\\
This work was supported by the Ministry of Science and Technology of the Republic of Croatia.


\end{document}